\documentstyle[12pt,doublespace]{article}

\thicklines                         
\begin{document}
\pagestyle{empty}

\begin{flushright}
CERN-TH.7221/94
\end{flushright}

\hrule width 0pt
\vspace{2.0cm}

\singlespace
\begin{center}
{\Large{\bf SEARCH FOR THE}}\\
{\Large{\bf INVISIBLY DECAYING HIGGS PARTICLE}}\\
{\Large{\bf AT LEP AND THE LHC}}

\bigskip

{\large D.P. ROY} \\ Theory Division, CERN, CH--1211 Geneva 23,
Switzerland\\

and

Theoretical Physics Group, Tata Institute of Fundamental
Research, Bombay 400 005, India\footnote{Permanent address}

\end{center}

\vspace{5cm}

\begin{abstract}
The Higgs particle can have dominantly invisible decay in a large
class of Majoron models as well as some SUSY models.  The LEP
signal and mass limit for a Higgs particle undergoing invisible
decay are explored.  They are found to be very similar to those for
the standard model decay.  The best signatures and discovery limit for
an invisibly decaying Higgs particle at the LHC are also discussed.
\end {abstract}

\vfill

\begin{flushleft}
CERN-TH.7221/94\\
April 1994
\end{flushleft}

\noindent (Invited Talk at the XXIXth Rencontres de
Moriond on QCD and High Energy Hadronic Interactions, M\'eribel,
France, 19--26 March 1994.)
\pagestyle{empty}
\clearpage\mbox{}\clearpage
\setcounter{page}{1}
\pagestyle{plain}
\newpage
\onehalfspace

The signatures for Higgs particle detection at LEP and the future
hadron colliders (LHC/SSC) have been extensively studied in the
framework of the standard model (SM) and its supersymmetric (SUSY)
extensions$^{1,2)}$. There exist some extensions of the SM, however,
with a qualitatively different signature for the Higgs particle. These
extensions are generically called the Majoron models (MM)$^{3-7)}$
and have been quite popular, e.g. in the context of generating neutrino
mass. They are characterized by the existence of a Goldstone boson
(the Majoron). Since the coupling of this Goldstone boson to the Higgs
particle is not required to be small on any theoretical or
phenomenological grounds, the Higgs particle could decay into an
invisible channel containing a Majoron pair$^{7-9)}$. Indeed the
importance of extending the Higgs search to this invisible decay
channel has been repeatedly emphasized over the past decade$^{8,9)}$.
However, quantitative investigations along this line have only
started very recently$^{10-12)}$.

The key features shared by essentially all Majoron models is a
spontaneously broken global $U(1)$ symmetry and a complex $SU(2)
\times U(1)$ singlet scalar field $\eta$ transforming non-trivially
under the global $U(1)$. The spontaneous breaking of the global $U(1)$
generates a massless Goldstone boson, the Majoron $J \equiv$~Im~$
\eta/\sqrt 2$, and a massive scalar $\eta_R \equiv \mathrm{Re}~
\eta/\sqrt
2$. The latter mixes with the massive neutral component $\phi_R$ of
the standard Higgs doublet through a quartic term $\phi^\dagger \phi
\eta^\dagger \eta$ in the scalar potential. Thus one has two massive
physical scalars
$$
H = \cos \theta \phi_R + \sin \theta \eta_R \quad \mathrm{and} \quad S
= \cos \theta \eta_R - \sin \theta \phi_R ~,
\eqno (1)
$$ where the mixing angle can always be chosen to lie in the range
0--45$^\circ$, so that the $H$ and $S$ have dominant doublet and
singlet components respectively. The above quartic term also generates
the following couplings of $H$ and $S$ to the massless Goldstone boson
$J$ : $$ {\cal L} = {(\sqrt 2 G_F)^{1/2} \over 2} \tan \beta
\left[M_S^2 \cos \theta S J^2 - M_H^2 \sin \theta H J^2 \right]~,
\eqno (2)
$$ where $\tan \beta = \langle\phi\rangle/\langle\eta\rangle$
 is the ratio of the two vacuum
expectation values$^{8)}$. The resulting decay widths of $H, S$ into
the invisible channel $(JJ)$ relative to the dominant SM channel $(b
\bar b)$ are
$$
\Gamma_{H \rightarrow JJ}/\Gamma_{H \rightarrow b \bar b} \simeq {1
\over
12} \left({M_H \over m_b}\right)^2 \tan^2 \theta \tan^2 \beta \left(1 -
{4m^2_b \over M_H^2} \right)^{-3/2}~,
\eqno (3)
$$
$$
\Gamma_{S \rightarrow JJ}/\Gamma_{S \rightarrow b \bar b} \simeq {1
\over
12} \left({M_S \over m_b}\right)^2 \cot^2 \theta \tan^2 \beta \left(1 -
{4m^2_b \over M_S^2} \right)^{-3/2}~.
\eqno (4)
$$
\vfill\eject
The large mass ratio $(M_{H,S}/m_b)^2$ on the r.h.s. implies that the
invisible decay channel could dominate for $S$ as well as $H$ over a
large range of the parameters $\tan \theta$ and $\tan
\beta$.\footnote{Both parameters depend on the scale of the global
$U(1)$ breaking relative to the $SU(2) \times U(1)$ breaking scale, on
which there are no severe phenomenological constraints.}

Although eqs. (1)--(4) above were derived for the simplest model$^{3)}$
having 1 singlet and 1 doublet scalar fields, similar considerations
hold for those having a larger Higgs content$^{5-8)}$ or a larger global
symmetry group than $U(1)$$^{9)}$. It may be added here that the Higgs
particles can also decay invisibly in the SUSY models via a pair of
lightest superparticles (LSP). For the minimal supersymmetric standard
model (MSSM) this invisible decay mode has been shown to dominate only
over a tiny range of parameters for the scalar Higgs particles but
over a larger range for the pseudoscalar$^{13)}$.

Thus it is important to extend the Higgs search strategies
at LEP and the LHC
to cover the possibility of a dominantly invisible decay. This is
simple at
LEP, since the dominant channel for the Higgs search is the
same for the SM
and the invisible decays -- i.e. the missing energy channel with one or
two jets$^{10-12)}$. It corresponds to the Bjorken production process
$$
e^+ e^- \buildrel Z \over \rightarrow Z^\ast H
\eqno (5)
$$
followed by $Z^\ast \rightarrow \nu \bar \nu ~, ~H \rightarrow b \bar b$
for the SM decay and $Z^\ast \rightarrow q \bar q ~,~ H \rightarrow JJ$
for the invisible decay. Indeed the larger branching fraction of
$Z^\ast$
into quarks implies a larger event rate for the latter case.  Figure 1
shows the expected number of signal events for the two cases$^{11)}$
for the ALEPH data sample of ref. [2].  The cuts reduce the signals by
only a factor of $\sim 2/3$ in either case while completely
eliminating the background.  Thus the signal size for the invisible
Higgs decay is a factor of 2--3 higher than the SM decay; and the
corresponding 95\% CL mass limit is higher by $\sim 6$ GeV.  This
would imply an $H$ mass bound somewhere between these two limits,
depending on the relative size of the two decay models (eq. (3)).  On
the other hand the production cross-section (5) would be suppressed by
a factor of $\cos^2 \theta~(\geq 0.5)$ since the $Z$ couples only to
the doublet component of the Higgs field.  Combining the two effects
leads to a $M_H$ bound in the Majoron models, which is within $\pm 6$
GeV of the SM value, irrespective of the model parameters, i.e. $48
\pm 6$ GeV~$^{2,11)}$ going up to $60 \pm 6$ GeV with the new ALEPH
data$^{12)}$.  A similar correlation between the Higgs signatures and
discovery limits for the two models is expected to hold for LEP II
as well.  It may be noted that the dominantly
singlet Higgs $S$\break\hfill
\newpage
\noindent
of the MM can be arbitrarily light for sufficiently small mixing angle
$\theta$, since the production cross-section is suppressed by a factor
of $\sin^2\theta$.

At the LHC, the missing energy is not measurable in view of the large
energy loss along the beam pipe.  So one has to convert it into a
missing-$p_T~(p\!\!/_T)$ signature by looking at one of the following
associated production processes at large $p_T$:
$$
(i)~H + {\rm jet}, \;\;\; (ii)~H + Z, \;\;\; (iii)~H+W \;\;
\mathrm{and}
\; (iv)~H + t\bar t~,
\eqno (6)
$$
followed by the invisible decay of $H$.  While the first process has too
large a background from $Z (\longrightarrow \nu\bar \nu)
+~\mathrm{jet}$,
the second and third processes are expected to give viable signatures at
the LHC$^{14)}$.  Figure 2 shows the Higgs signal from $(ii)$
 along with the
dominant background in the $\ell^+\ell^-p\!\!/_T$ channel$^{14)}$.
For $p\!\!/_T > 200$ GeV, one gets
a viable signal size $(\sim 2.5~$fb) and signal/background ratio
($\sim 1/2)$.  Figure 3 shows the signal
from $(iii)$ along with the irreducible background in the $\ell
p\!\!/_T$
channel$^{14)}$.  Here the signal size is $\sim 5~$fb and the
signal/background ratio is $\sim 1$ for $p\!\!/_T > 200$ GeV.  Thus
one should be able to probe the intermediate mass range of Higgs
 (100--200 GeV) even after making allowance for the
 suppression factor of
$\cos^2 \theta$ at the production vertex.  For $M_H > 200$ GeV, the $H
\rightarrow WW, ZZ$ decay modes are expected to dominate over the
invisible $(JJ)$ mode.

The fourth process has also been shown to give a signal/background ratio
of $\sim 1$ at the LHC$^{16)}$.  But the signal size is relatively small
$(\sim 0.5~$fb).  Besides, this signal is far more demanding on the
detector performance, as it requires good $b$ identification as well as
reconstruction of $W$ and $t$ masses from hadronic jets.

Finally, it may be noted that the signals of Figs. 1--3 are equally
applicable to the invisible decay of Higgs scalar into a pair of LSP
in the SUSY models.  In fact they should be exact in this case since
there is no singlet Higgs scalar $S$, i.e. $\cos \theta = 1$.  However
they do not apply to the pseudoscalar Higgs boson $A$ of SUSY models,
since it does not couple to gauge bosons.  Only the last process
mentioned above is applicable to the invisible decay of $A$~$^{16)}$.

\newpage

\singlespace
\noindent\underbar{\bf References}

\begin {enumerate}

\item
See e.g. J.F. Gunion, H. Haber, G.L. Kane and S. Dawson, The Higgs
Hunter's Guide (Addison-Wesley, Reading, MA, 1990).

\item
ALEPH Collaboration : D. Decamp et al., Phys. Rep. {\bf 216} (1992) 253.

\item
Y. Chikashige, R.N. Mohapatra and R.D. Peccei, Phys. Lett. {\bf 98B}
(1980) 265.

\item
D.B. Reiss, Phys. Lett. {\bf 115B} (1982) 217; F. Wilczek, Phys. Rev.
Lett. {\bf 49} (1982) 1549.

\item
G. Gelmini and M. Roncadelli, Phys. Lett. {\bf B99} (1981) 411.

\item
A.S. Joshipura and S.D. Rindani, Phys. Rev. {\bf D46} (1992) 3000;
\hfill\break
A.S. Joshipura and J.W.F. Valle, Nucl. Phys. {\bf B397} (1993) 105.

\item
J.C. Romao, F. de Campos and J.W.F. Valle, Phys. Lett. {\bf B292} (1992)
   329.

\item
R.E. Schrock and M. Suzuki, Phys. Lett. {\bf B110} (1982) 250;
\hfill\break
L.F. Li, Y. Liu and L. Wolfenstein, Phys. Lett. {\bf B159} (1985)
 45; \hfill\break
E.D. Carlson and L.B. Hall, Phys. Rev. {\bf D40} (1985) 3187;
\hfill\break
G. Jungman and M.A. Luty, Nucl. Phys. {\bf B361} (1991) 24; \hfill\break
A.S. Joshipura and S.D. Rindani, Phys. Rev. Lett. {\bf 69} (1992) 3269.

\item
J.D. Bjorken, Invited Talk at the Symposium on the SSC Laboratory,
Corpus Christi,
Texas, October 1991, SLAC-PUB-5673 (1991).

\item
A. Lopez-Fernandez, J.C. Romao, F. de Campos and J.W.F. Valle, Phys.
Lett. {\bf B312} (1993) 240.

\item
B. Brahmachari, A.S. Joshipura, S.D. Rindani, D.P. Roy and K. Sridhar,
Phys. Rev. {\bf D48} (1993) 4224.

\item
ALEPH Collaboration : D. Buskulic et al., Phys. Lett. {\bf B313} (1993)
312.

\item
A. Djouadi, J. Kalinowski and P.M. Zerwas, Z. Phys. {\bf C57} (1993)
569.

\item Debajyoti Choudhury and D.P. Roy, Phys. Lett. {\bf B322} (1994)
368.

\item
M. Diemoz, F. Ferroni, E. Longo and G. Martinelli, Z. Phys. {\bf C39}
(1988) 21.

\item
J.F. Gunion, Phys. Rev. Lett. {\bf 72} (1994) 199.

\end {enumerate}
\newpage
\noindent\underbar{\bf Figure captions}
\begin{itemize}
\item[Fig. 1] The expected Higgs signals for the SM and MM
(invisible) decay modes corresponding to the published ALEPH
data$^{2)}$.
\item[Fig. 2] The $HZ$ signal (dotted and dashed lines) and
the $ZZ$ background (solid line) cross-sections for the
dilepton + missing-$p_T$ channel at the LHC, calculated
using the DFLM structure functions$^{15)}$.
\item[Fig. 3] The $HW$ signal (dotted and dashed lines)
and the $WZ$ background (solid line) cross-sections for
the lepton + missing-$p_T$ channel at the LHC, calculated
using the DFLM structure functions$^{15)}$.
\end{itemize}

\end {document}